# Molecular dynamics simulations and in vitro analysis of the CRMP2 thiol switch

Daniel Möller[+1], Manuela Gellert[+2], Walter Langel[1], Christopher Horst Lillig[*2]

[+] These authors contributed equally to this work.

[*] corresponding author: horst@lillig.de

[1] Biophysical Chemistry, Institute of Biochemistry, University Greifswald, 17489 Greifswald, Germany

[2] Institute for Medical Biochemistry and Molecular Biology, University Medicine, University Greifswald, 17489 Greifswald, Germany

## Abstract

Collapsin response mediator protein CRMP2 (gene: DPYSL2) is crucial for neuronal development. The homotetrameric CRMP2 complex is regulated via two mechanisms, first by phosphorylation at, and second by reduction and oxidation of the Cys504 residues of two adjacent subunits. Here, we analyzed the effects of this redox switch on the protein *in vitro* combined with force field molecular dynamics (MD). Earlier X-ray data contain the structure of the rigid body of the molecule but lack the flexible C-terminus with the important sites for phosphorylation and redox regulation. An *in silico* model for this part was established by replica exchange simulations and homology modelling, which is consistent with results gained from CD spectroscopy with recombinant protein. Thermofluor data indicated that the protein aggregates at bivalent ion concentrations below 200 mM. In simulations the protein surface was covered at these conditions by large amounts of ions, which most likely prevent aggregation. A tryptophan residue (Trp295) in close proximity to the forming disulfide allowed the measurement of the structural relaxation of the rigid body upon reduction by fluorescent quenching. We were also able to determine the second order rate constant of CRMP2 oxidation by $H_2O_2$. The simulated solvent accessible surface of the hydroxyl group of Ser518 significantly increased upon reduction of the disulfide bond. Our results give first detailed insight in the profound structural changes of the tetrameric CRMP2 due to oxidation and indicate a tightly connected regulation by phosphorylation and redox modification.

# Introduction

CRMP2 is an essential component of the semaphorin 3A signaling pathway and crucially involved in cell polarization and migration and thus indispensable for neuronal development and cell mobility[1,2]. This CRMP2 pathway controls, for instance, ureteric and vascular patterning[3], axon guidance[4], and the migration and polarization of T-cells[5,6]. The protein was implied in various physiological disorders, for instance a hyper phosphorylated form was found in protein deposits during Alzheimer's disease[7,8]. The regulation of CRMP2 and the biological effects triggered by this signaling hub are complex.

The crystal structures[9,10] and further analyses[11] demonstrated a homotetramer as the native conformation of CRMP2. The full sequence of a human CRMP2 monomer (Q16555) consists of 572 amino acids (aa). Unfortunately, none of the two crystal structures known to us cover the full protein, but only Asp15 - Ala489 for 2GSE[9] or Ser14 - Glu490 for 2VM8[10]. The steric structure is stabilized by four bivalent ions such as $Ca^{2+}$ or $Mg^{2+}$.

A thiol-disulfide switch[11,12] regulates the protein. Two Cys504 residues of two monomers form an inter-molecular disulfide bond[11]. Closing and opening of these bonds yield an oxidized and a reduced form of CRMP2, respectively. This dithiol-disulfide switch of CRMP2 profoundly affects its structure and function. In most cases only one of the two possible disulfide bonds in a tetrameric complex is formed. Our previous work suggested significant conformational changes in the homo-tetrameric CRMP2 complex upon oxidation, leading to a decreased exposure of hydrophobic aa at the surface[11]. The protein is regulated not only via the redox switch[11-13], but also by direct phosphorylation at multiple sites via protein kinases[14-16]. The Rho kinase ROCK phosphorylates CRMP2 at Thr555, cycline-dependent kinase 5 (CDK5) at Ser522. While these two steps are thought to be independent, interconversion of CRMP2 by glycogen synthase kinase 3 beta (GSK3β) at Thr509, Thr514, and Ser518 strictly requires the priming phosphorylation at Ser522 by CDK5[14,16]. The phosphorylation of CRMP2 controls its binding to other proteins, for instance tubulin[4], and the Cyfip1/WAVE1 (cytoplasmic FMR1-interacting protein 1/WASP family verprolin-homologous protein-1) complex[17], a regulator of actin branching and polymerization. All of these experimentally verified phosphorylation sites are localized within the flexible C-terminus of the protein downstream of the Cys504 redox switch.

According to Majava et al. and Stenmark et al., the CRMP2 homotetramer has to be stabilized by four bivalent ions such as $Ca^{2+}$ or $Mg^{2+}$, which are integrated into the molecular structure, and the removal of these ions will induce denaturation[9,10].

Additionally, the bivalent ions attach to the charged surface side chains, resulting in shifts in the electrostatic surface potentials and an increase of the surface areas around the C-termini. Under the conditions of the work of Majava et al., ionic concentrations around 20 mM had a stabilizing effect, whereas 200 mM or higher again resulted in protein unfolding[10].

In this study, we modelled and compared the completely oxidized and reduced states of CRMP2 and analyzed the changes in the secondary structure, the solvent accessible surface area (SASA) and the composition of defined aa properties at the surface. By tryptophan fluorescence, the transition between oxidized and reduced states of CRMP2 was probed. We further studied the stability of CRMP2 in ionic environments and the location of the bivalent ions on specific amino acids.

# Results

We hypothesized that CRMP2s regulation by oxidation and reduction leads to profound structural and conformational changes. These changes were investigated by analyses of CRMP2 *in silico* and *in vitro* using a variety of experimental methods and force field molecular dynamics.

## *Secondary structure and conformational changes between reduced and oxidized CRMP2 assayed by MD simulations and CD spectroscopy*

The steric structure of CRMP2 was analyzed using three methods. Results from CD spectroscopy (**Figure 1**) average over the full protein with 572 amino acids per monomer, including the flexible parts of the C- and N-termini with unknown secondary structures. Molecular dynamics simulations cover residues 16 to 531, and the X-ray results reproduce the rigid part of the protein only (14-490). Taking into account these differences, experimental data and simulations yield consistent distributions of the secondary structure motifs.

Both experimental crystal structures of CRMP2, 2VM8[10] and 2GSE[9] are rather similar with 56.5 % helical structure (α-helices and turns) as largest structured part (**Table 1**). The MD calculations are based on these X-ray data, and yield a major extent of helical structure elements.

Helices and turns are very similar, and it is not possible to distinguish between them in CD spectra. The total contribution of these two motifs increases from 57 % to 62 % upon reduction. Similarly to CD, the simulation qualitatively also yields an enhancement of helical structures in the reduced conformation of the protein, but the effect is rather small, increasing from 57.6 % to about 58.5 %.

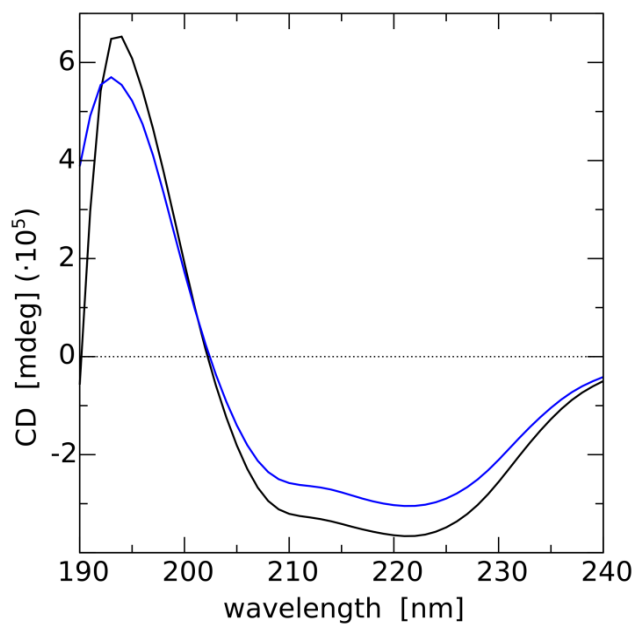

**Figure 1: CD spectra of oxidized (black) and reduced (blue) recombinantly expressed CRMP2, at 10 μM in 20 mM Mg$^{2+}$ solution. The traces clearly indicate a shift in the overall secondary structure upon oxidation. We estimated the secondary structure contents of the spectra by the method of Raussens et al.[18].**

The β-sheets contribute 23 % to the crystal structure, which decreases to a constant value of 20 % in the MD simulations in solution. CD spectroscopy revealed even lower values of only 10-13 % and significant parts of the structure are assigned to coil (28 %). *In silico*, about 20 % are random coil, and a small shift from this disordered to a helical structure is observed when switching from the oxidized to the reduced protein conformation.

The comparison between crystal, *in silico*, and the data from CD spectroscopy suggests that the C-terminus, which is missing in the crystal structure, mainly adds helical elements to the *in silico* and *in vitro* protein. Remaining amino acids are unstructured or helical, resulting in a decrease of the relative amount of β-sheet and an increase of the helical proportion. No correlation was detected between secondary structure and ionic concentrations in the MD.

## *Exposure of surfaces to solvent*

The traces for the SASA in **Figure 2** show a steep increase within 50 ns indicating structural relaxation. After this time, the values fluctuate around a constant value. We thus assume that the simulation times of 200 to 300 ns were sufficient for sampling equilibrated systems. The highest SASA is found for the protein molecules with the four stabilizing bivalent ions (blue and black traces). The run without bivalent ions shows a large fluctuation, whereas the systems with higher ion concentrations have smaller areas and may be shrinking under the influence of the surface charge, yielding more compact protein structures. The results for the radius of gyration (Supporting Information **Figure SI 3**) of these systems seem to confirm this picture. Further inspection of the data demonstrates that the fluctuation or 'breathing' of the relaxed surface is mainly due to a fluctuation of the SASA of the flexible C-terminal area.

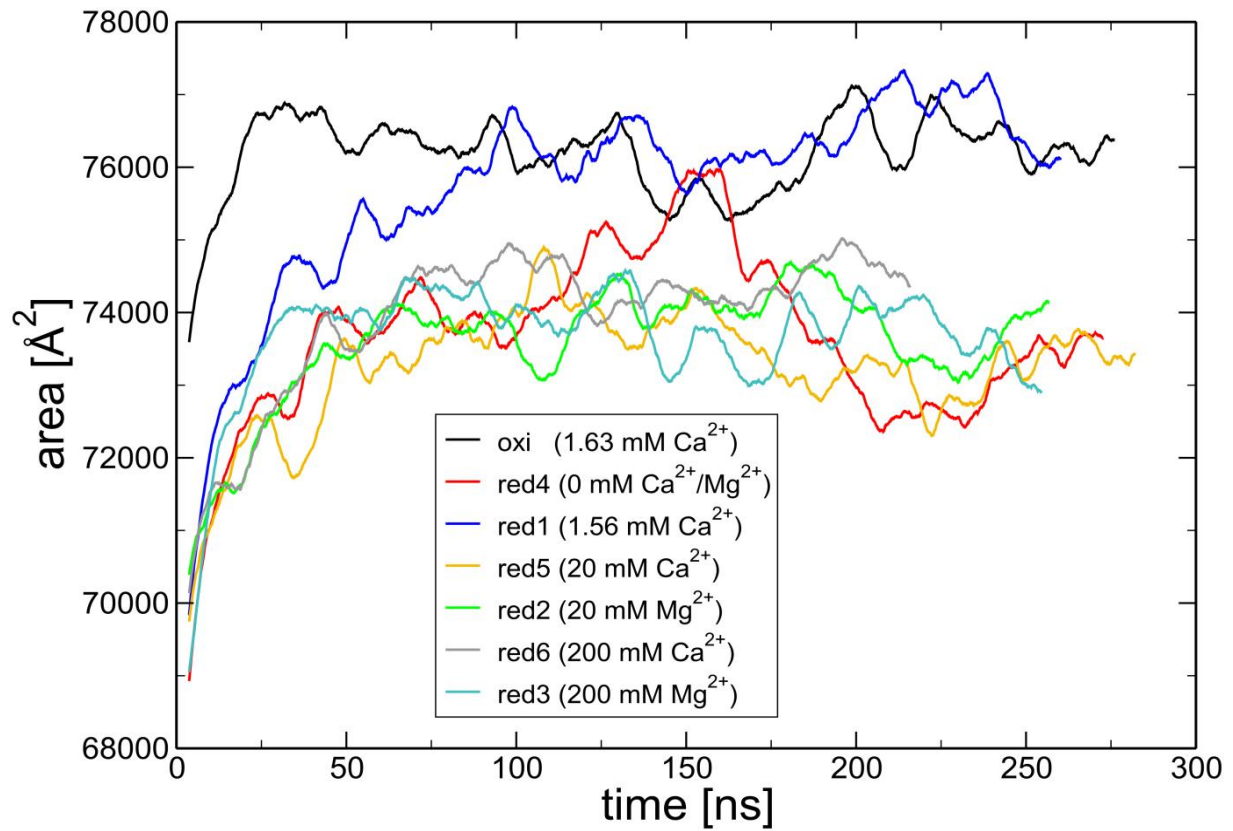

Figure 2: Solvent accessible surface area (SASA) of the CRMP2 homotetramer as a function of time from the MD simulations: Oxidized with four structure stabilizing bivalent ions (black), reduced without any bivalent ions (red), with four structure stabilizing bivalent ions (blue), with 20 mM ($Mg^{2+}$ green, $Ca^{2+}$ yellow) and with 200 mM bivalent ions ($Mg^{2+}$ cyan, $Ca^{2+}$ grey). The curves are running averages of the raw data with a window of 7.5 ns.

We consider the amino acids Ala, Ile, Leu, Met, Phe, Pro, Trp, Val as hydrophobic and evaluate the ratio of their solvent exposed area to the total SASA. During reduction and with enhanced the ion concentration from 0 to 200 mM, this ratio increased by less than 1 % (**Table 1**).

## SASA of the CRMP2 phosphorylation sites Thr509, Thr514, Ser517, Ser518 and Ser522

In a previous study, we used HeLa cells expressing Grx2c, which leads to the complete reduction of the normally oxidized CRMP2 *in vivo* [REF: Gellert et al., Cancer Research submitted]. In these cells, an increased CRMP2 phosphorylation at Thr509, Thr514, and Ser518 was described. These three residues are the target sites for phosphorylation by the kinase GSK3β. For a side chain, which may be phosphorylated *in vivo*, the simulation should yield a structure, in which the hydroxyl group is well accessible.

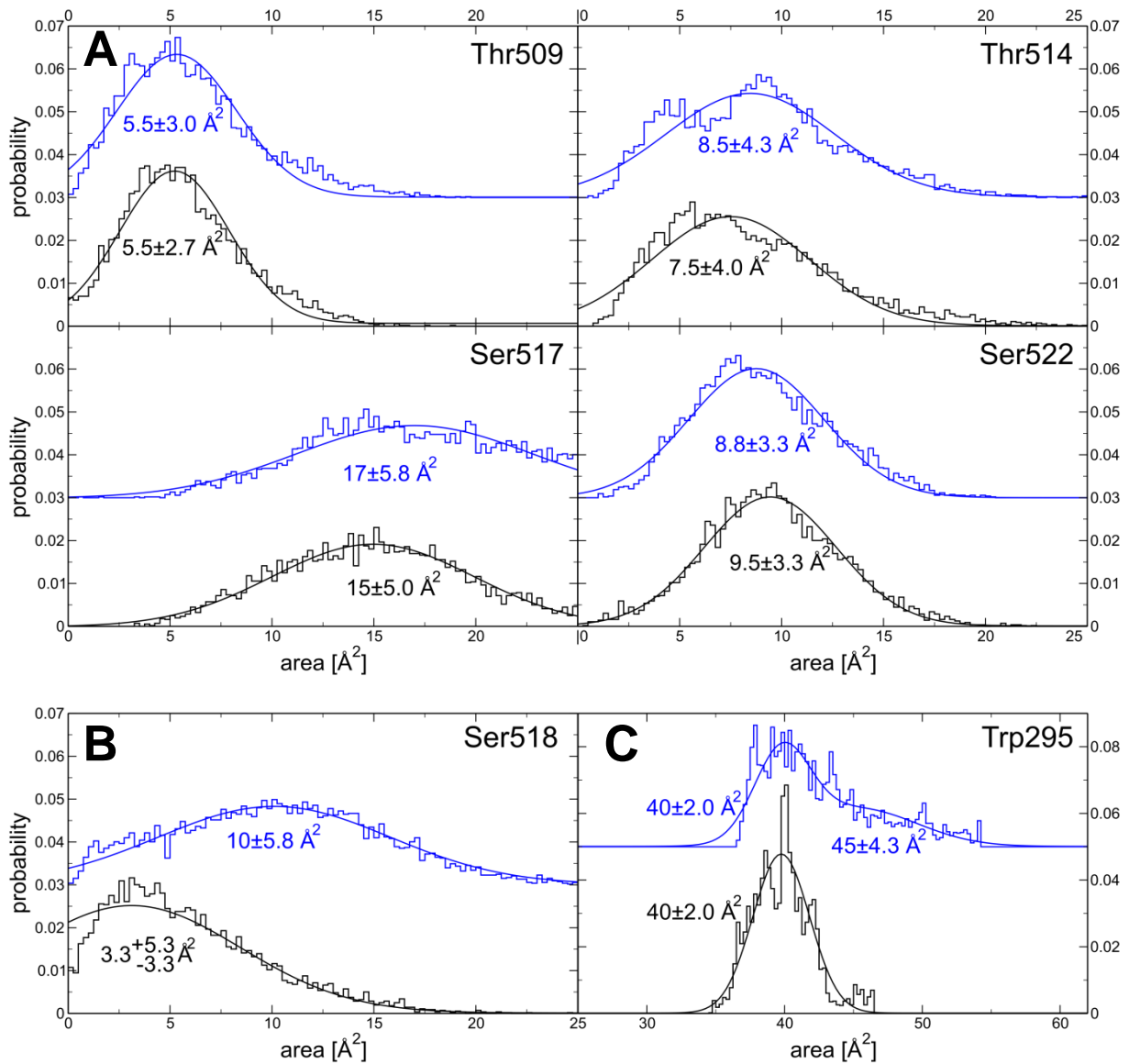

**Figure 3: Distributions of the SASA:**

(A) **hydroxyl groups of CRMP2s phosphorylation sites at the C-terminus in MD simulations. The molecules contained four bivalent ions. All values show significant fluctuations due to thermal motion. Centers and widths of the Gaussian fits are indicated in the figures. The distributions for oxidized (black) and reduced CRMP2 (blue) are not significantly different for Thr509, Thr514, Ser517 and Ser522.**

(B) **Only the hydroxyl group of Ser518 is significantly more exposed to the solvent in the reduced conformation compared to the oxidized form.**

(C) **The distribution of the full Trp295 SASA in the oxidized state is well described by a Gaussian centered at 40 Å². In the reduced system, a second Gaussian at 45 Å² with about the same intensity has to be added indicating an increase of the average SASA due to the relaxation of the protein.**

In our model, the five side chains are all in contact with the solvent (**Figure 3**). The hydroxyl group of Ser517 has a significantly higher SASA than the other four amino acids. Four of the hydroxyl groups do not show a significant response to reduction and protein relaxation. However the SASA of Ser518 significantly decreased after oxidation of Cys504, suggesting that this serine loses contact to the solution and therefore GSK3β accessibility. The variation of bivalent ion concentrations did not significantly affect the SASA data.

It may depend on the structure of the respective enzyme, if phosphorylation of serine affords solvent exposure of its hydroxyl group only, or of the full side chain. The inspection of the SASA for the full amino acids shows that Ser522 seems to be buried in all systems and is less exposed to the solvent than the other four amino acids. We speculate that this is consistent with the previously described observation that Ser522 is phosphorylated by another kinase, CDK5, than the other residues[14,16].

## *Oxidation of CRMP2 by $H_2O_2$ in vitro and Quenching of Tryptophan Fluorescence*

The Trp295 is part of the rigid body of CRMP2, but is located in close proximity to the flexible C-terminus (**Figure 4.A**). We assume that the observed changes in tryptophan fluorescence (**Figure 4.B**) are mainly based on the changes in the environment of Trp295. During oxidation of the recombinant protein by $H_2O_2$ *in vitro*, the tryptophan fluorescence (**Figure 4.B**) was quenched. Obviously, the environment of the Trp295 residue in the CRMP2 body was affected by the oxidation reaction in the flexible C-terminus resulting in fluorescence quenching. The MD showed that because of this rearranging the SASA of this residue is slightly smaller after the formation of the disulfide bridge (**Figure 3**). *In vitro* we detected a continuous red shift of the emission maxima during the oxidation, also indicating changes of Trp295s environment (Insert of **Figure 4.B**). The displacement from 340.85 to 343.10 nm corresponded to a decrease of the quantum energy by 2.3 kJ/mol (**Figure 4.C**). It is well known that tryptophan fluorescence depends on the environment of the residue. Therefore we assign this red shift to a shift of the energy levels of the fluorescent indole ring of the corresponding Trp295 (supplementary **Figure SI 4**).

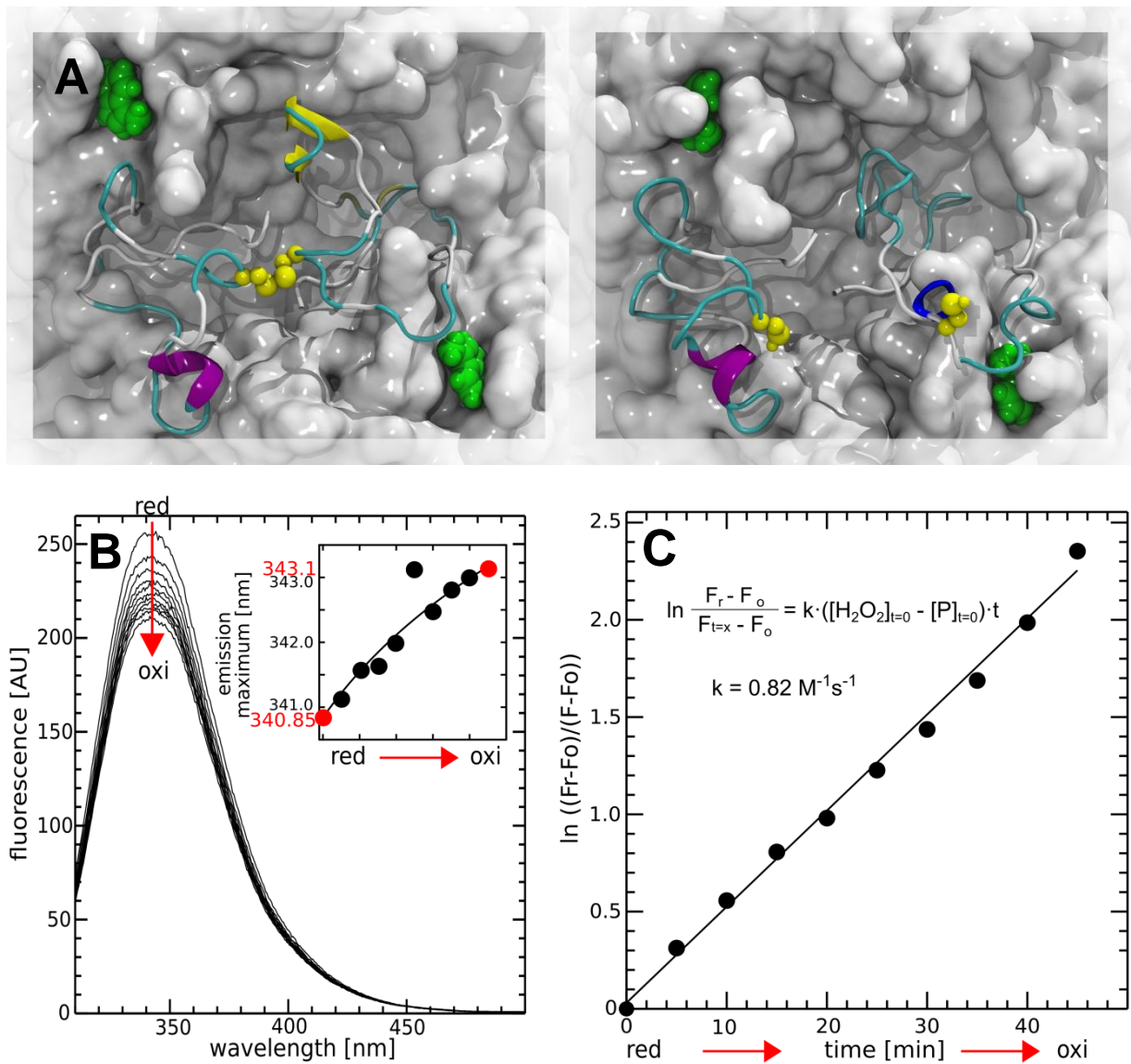

**Figure 4: Fluorescence quenching of the Trp295.**

(A) Snapshots of the structures of Trp295 (green) on the surface of oxidized (right) and reduced CRMP2 (left) from MD. The 41 aa of C-termini are shown with their secondary structure (helix: purple and blue; sheet: yellow arrows; turn: cyan; coil: white) and the Cys504 residues as ball-and-stick in yellow.

(B) Fluorescence spectra of CRMP2 after excitation at 296 nm, measured every 5 min. The intensity continuously decreases during the oxidation with $H_2O_2$. The insert shows the red shift of the emission maxima over time.

(C) Evaluation of the decrease of the emission at 340 nm according to the inserted equation. The slope of the straight line corresponds to a pseudo first order rate constant of $8.2 \cdot 10^{-4} \text{ s}^{-1}$. With the concentration of $10^{-3}$ mol/l for $H_2O_2$, we obtained a second order rate constant of $0.82 \pm 0.06 \text{ M}^{-1} \text{ s}^{-1}$ (n=3).

## Correlation of Trp295 fluorescence wavelength and electrostatic energies of CRMP2s oxidized and reduced state

In the frame of a force field model, the shift of an electronic state relative to vacuum energy is given by the sum of the products of the atomic point charges $q_a$ of the atoms $a$ and the electrostatic potential $V_a$ of the atomic charges of the surrounding molecules[19]:

$$E = e * (\sum_a V_a * q_a) \tag{1}$$

The smooth particle-mesh Ewald method (PME) in VMD yielded a grid for the electrostatic potential with a resolution of 0.5 Å and an Ewald factor of 0.25, which was interpolated at the positions of the tryptophan atoms a yielding $V_a$. This analysis afforded the knowledge of the partial charges $q_a$ and positions of the atoms a, which were only known for the electrostatic ground state. These energies E thus only refer to this ground state.

The analysis of the E from the Trp295 side chain (**Figure SI 4**) yielded two results. By oxidizing the molecule, the electrostatic energy of the ground state of the indole ring increased by 1.1 kJ/mol, which may contribute to the observed red shift of the emission maxima. We further analyzed the electrostatic energy of indole and its benzene and pyrrole components as a function of the ionic strength of the solution. The addition of ions to the solution provoked a shift of the electrostatic energies of the ground state of Trp295. The overall electrostatic energy of the indole ring is nearly independent of ion type or concentrations, but benzene and pyrrole rings responded in opposite directions to the ions. The energy of the benzene part increases with increasing $Ca^{2+}$ concentration, but decreases with increasing $Mg^{2+}$ concentration. In contrast, the pyrrole part shows an increase of energy with $Mg^{2+}$ and a decrease with $Ca^{2+}$.

## Ion-interactions and Influences of divalent cations on the thermal stability of CRMP2

We analyzed the effects of the concentration of divalent cations, *i.e.* $Mg^{2+}$ or $Ca^{2+}$, on the thermal stability of the recombinant protein in a thermofluor assay[20]. Re-buffering of the protein in 10 mM HEPES buffer (2-[4-(2-hydroxyethyl)piperazin-1-yl]ethane sulfonic acid; pH 8.0) using gel filtration chromatography, resulted in precipitation. This could only be prevented by the immediate addition of 200 mM $Mg^{2+}$ ions. We thus eluted the protein in HEPES buffers containing either 300 mM $Mg^{2+}$ or $Ca^{2+}$ ions, which prohibited the precipitation completely. The ion concentrations for the thermofluor assay were then adjusted by addition of appropriate concentrations of the EDTA, which efficiently chelates divalent cations.

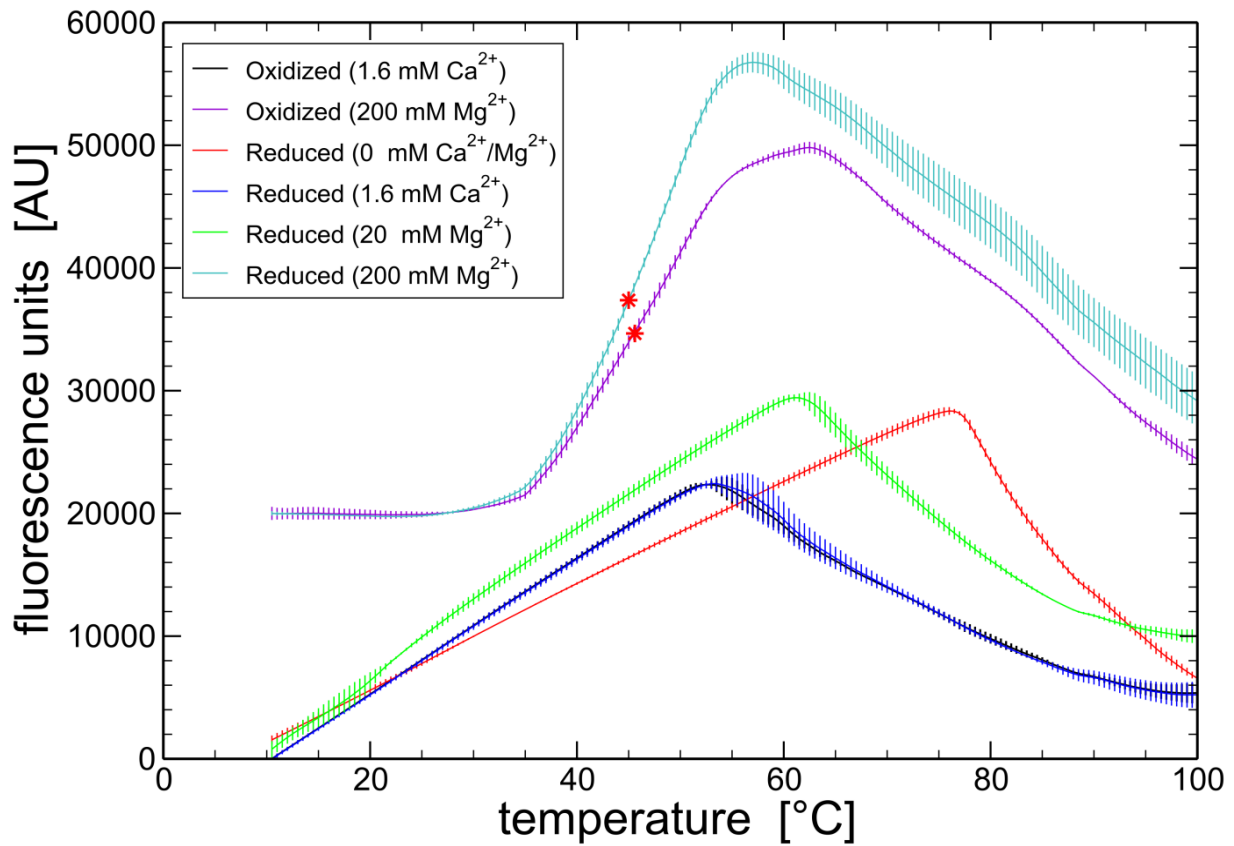

**Figure 5: Differential scanning fluorimetry of recombinant CRMP2 at different ion concentrations. The protein was kept at 300 mM ion concentration, before EDTA was added to set the conditions to the ion concentrations indicated. The standard deviation was included with the curves (n=3). The curves of oxidized and reduced CRMP2 at 200 mM Mg$^{2+}$ (violet and cyan) were shifted by 20000 au.**

The fluorescence curves of all samples showed a strong increase in the temperature range from 10-60 °C (**Figure 5**). The data recorded with 200 mM Mg$^{2+}$ were fitted by the standard Boltzmann model[21] for the intensity I at the temperature T:

$$I(T) = I_L + \frac{I_H - I_L}{1 + \exp\left(\dfrac{T_m - T}{I'}\right)} \qquad (2).$$

$I_H$ and $I_L$ are the respective maximum and minimum fluorescence intensities and $I'$ is related to the maximum slope. Both reduced and oxidized CRMP2 are obviously stable under these conditions with melting temperatures of $T_m$ = 45 and 45.6 °C (red stars in **Figure 5**), respectively. At lower Ca$^{2+}$ or Mg$^{2+}$ concentrations, the intensity nearly linearly increased over a larger temperature range, likely due to precipitation of the proteins.

It may be possible that the degradation of the samples at low ion concentrations in this experiment are not due to spontaneous denaturation of single homotetramers but due to their aggregation, which was beyond the scope of the MD simulation. Onset of protein denaturation should be indicated in the simulations by an increase of the hydrophobic area exposed to the solvent (**Table 1**). Thus was less than 1 %, and did not suggest the induction of CRMP2s denaturation.

## *Cation distribution on CRMP2*

For each of the bivalent ions $Ca^{2+}$ and $Mg^{2+}$ we distinguish two layers around CRMP2 (**Figure 6.A**). At lower concentrations one very well defined shell is found, which contains ions interacting directly with the amino acid side chains and probably may be assigned as rigid Helmholtz plane. The second layer is much less occupied and has a broader distance distribution than the first one, possibly being more flexible. $Mg^{2+}$ has a smaller ion radius than $Ca^{2+}$ and has a shorter distance to the amino acids.

In the frame of the force field model, the bivalent ions interact mainly with the negative oxygen atoms on the protein surface. The predominant interaction centers are the deprotonated carboxyl groups of Asp and Glu (**Figure 6.B, C**). Especially at low ion concentrations, $Ca^{2+}$ ions also attach to the oxygen atom in the Gln side chain. In the data for the first two shells, some ions are found close to amino acids which do not contain negative centers in the side chains such as Arg, Ile, Leu, Pro and Thr. Inspection of the steric structures reveals unspecific interactions of the cations with the backbone carboxyl groups (**Figure 6.C**).

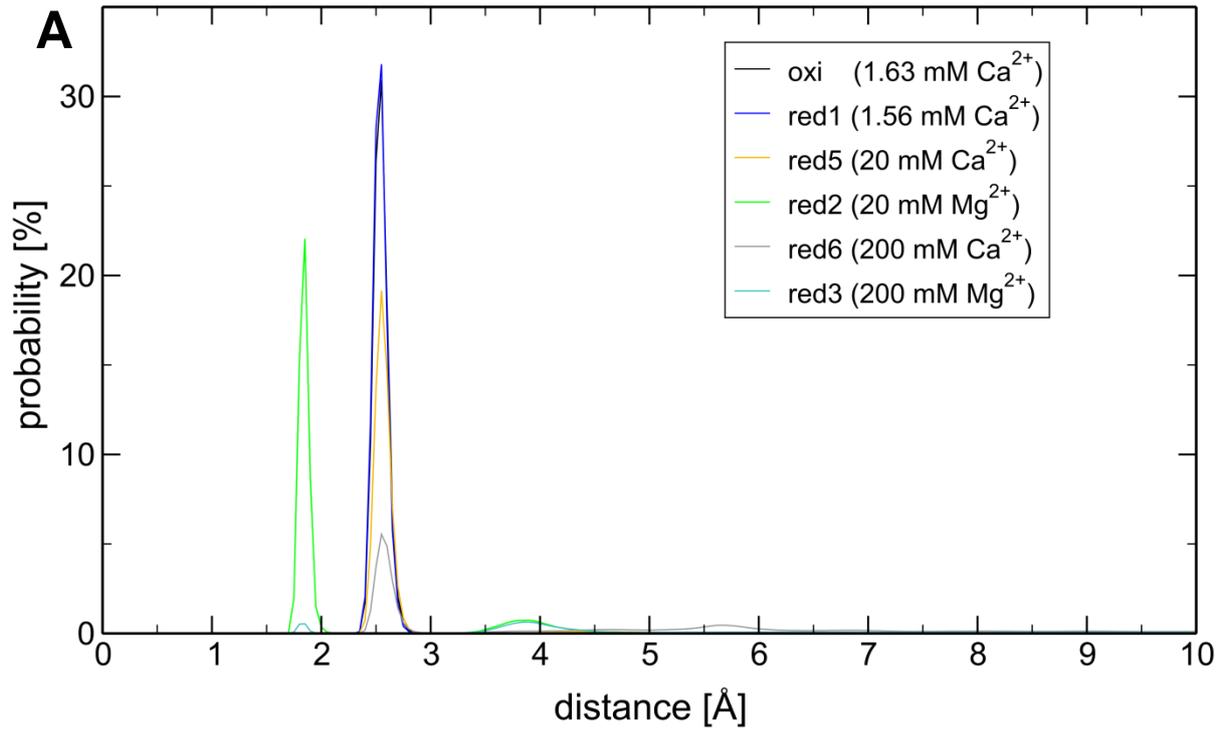

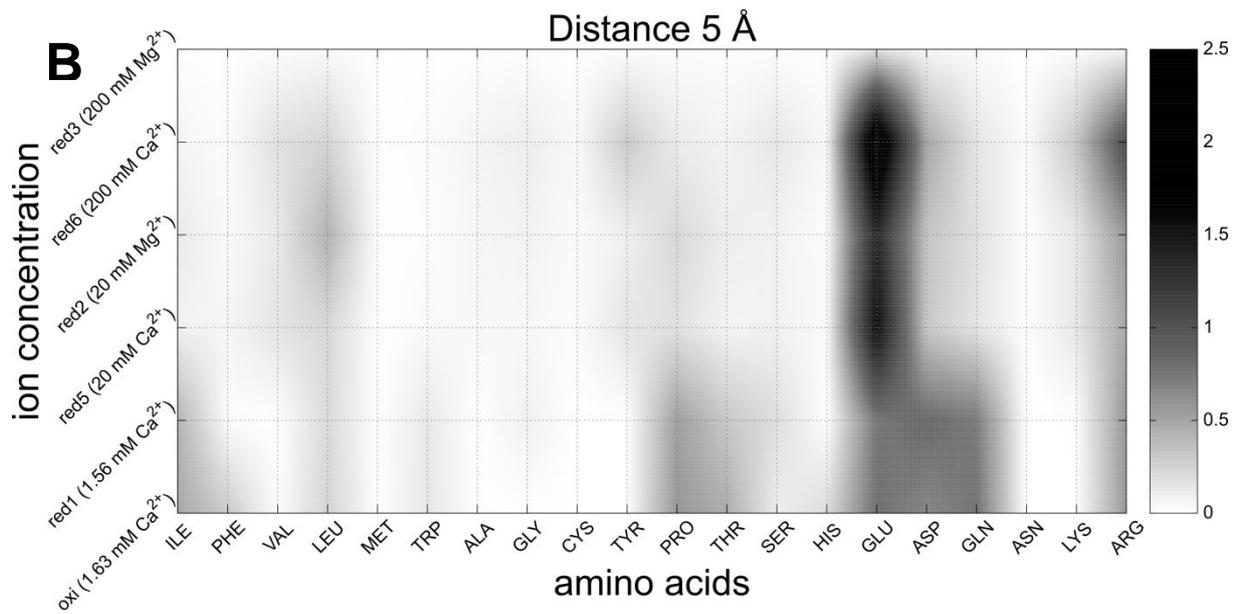

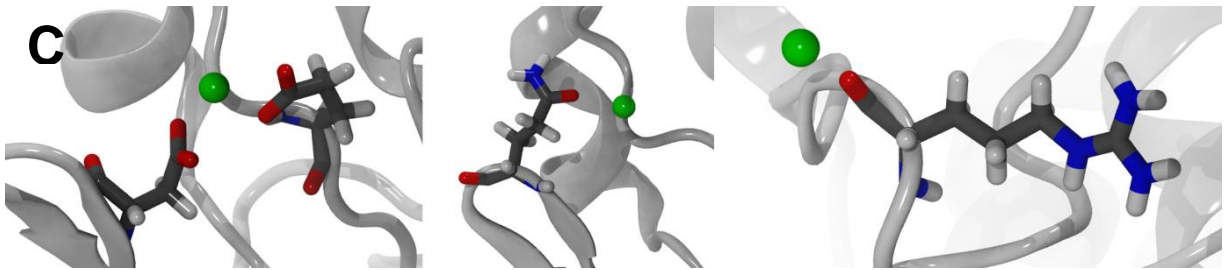

**Figure 6: Bivalent cations interacting with CRMP2.**

(A) **Normalized distribution of cation distances to the protein. The first well defined shells are clearly distinguished for $Mg^{2+}$ at 1.8 Å (green) and $Ca^{2+}$ at 3 Å (orange). At 200 mM concentration broader second shells are found with respective distances of 3.8 to 4 Å for $Mg^{2+}$ and for $Ca^{2+}$.**

(B) **Probability of the interaction of an amino acid with a bivalent ion up to distances of 5 Å. Data are normalized with respect to the number of occurrences of the respective amino acid at water exposed surface, the number of free ions and the simulation time. The amino acids are arranged according to their hydropathy[22,23]. Asp, Asn, Glu and Gln are sorted in order of functionality.**

(C) **Snapshots of the interactions between bivalent ions and amino acids. From left to right: Predominant attachment to the highly negative carboxyl groups of Asp and Glu, weaker interaction with the oxygen in the Gln side chain and weak adhesion to the carboxyl groups of several amino acids.**

## Discussion

### *Secondary structure*

The structure of CRMP2 used in this work is based on X-ray data. However they do not cover the chemical most important part of the protein, the C-terminus containing the regulatory phosphorylation sites and the disulfide forming cysteine residues. The C-terminus revealed to be very flexible and probably could not attain an ordered structure in the crystal. On the other hand, the regulatory sites must show high flexibility for structural rearrangements. We assembled a structure using molecular dynamics, which is consistent with experimental data and permitted further analysis of the protein. SASA data show that the fluctuation (also described as breathing) of the protein surface as well as the radius of gyration is mainly due to the flexible C-terminus rather than to the rigid body.

### *Tryptophan fluorescence*

However, also parts other than the C-termini are affected by the structural rearrangement of the molecule after oxidation. The SASA of the Trp295 decreases upon oxidation, resulting in a quenched and red shifted fluorescence. Our simulations reveal an increase of the electrostatic energy of the ground state of about 1.1 kJ/mol. The observed red shift of 2.3 kJ/mol is reproduced, assuming a down shift of the excited state by 1.2 kJ/mol (**Figure 7**). We have no estimation for the upper state, since the charge distribution in the upper state[24] is not accessible in our calculations. We do not know the fluorescing electronic state (La or Lb) [25,26], but the shift of the fluorescence quantum energy of Trp295 during oxidation is the same order of magnitude as the shift for the ground state from our force field data.

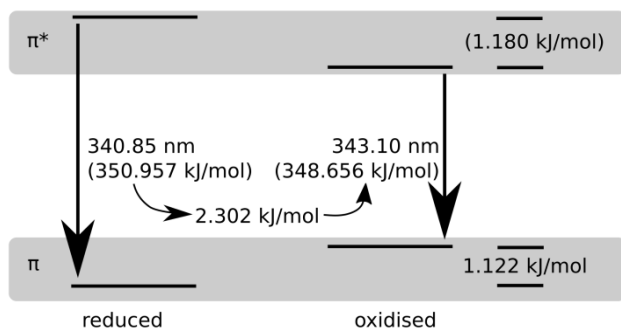

**Figure 7: Possible change in ground (π) and excitation state (π*) of the Trp295 fluorescence**

### *Ionic environment*

Our results demonstrated a more distinct response of CRMP2 to the bivalent $Mg^{2+}$ and $Ca^{2+}$ ions compared to the literature. We aimed to reproduce the results of Majava et al.[10] regarding the protein stability with two different bivalent ions ($Ca^{2+}$ or $Mg^{2+}$) in various concentrations. A stabilizing effect with 20 mM CaCl$_2$ or MgCl$_2$ was reported, but a concentration of 200 mM destabilized the protein significantly.

Our *in vitro* studies do not confirm these effects on CRMP2s structure, *e.g.* the postulated beginning of denaturation[10]. We saw a fast precipitation at lower ion concentrations up to 20 mM and, on the contrary, stabilizing effects at 200 mM. We propose is that the adsorption of these ions to the protein results in a positive zeta-potential, which is only high enough to stop aggregation at high ion concentrations in the solution.

## *Mechanism*

The molecular modeling of the CRMP2 thiol switch through our MD simulations suggested profound structural changes as a consequence of oxidation and reduction. These changes were confirmed *in vitro*. The fluorescence quenching of Trp295 could also be used to determine its redox state and to determine the rate constant of the oxidation of the protein by hydrogen peroxide. Moreover, our results suggest a complex cross talk between the redox switch and other post-translational mechanisms that regulate CRMP2s biological activity (**Figure 8**). We have previously demonstrated that all three target sites of the GSK3β kinase showed significantly increased levels of phosphorylation in cells in which CRMP2 was kept in the reduced conformation by Grx2c [REF: Gellert et al., Cancer Research submitted]. The molecular dynamics simulations as presented here further support and expand this hypothesis.

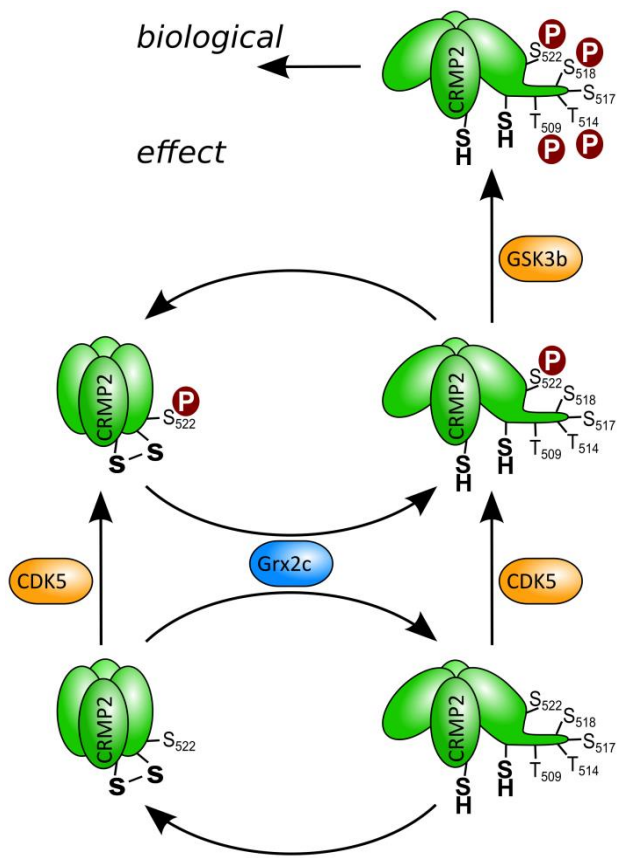

**Figure 8: Model of the posttranslational regulation of CRMP2's biological activity. Phosphorylation of Ser522 by the priming kinase CDK5 appears to be independent of the redox switch. However, the subsequent phosphorylation steps by the GSK3β kinase may require the reduction of the thiol switch first, followed by phosphorylation of Ser518 and its neighboring residues, i.e. Thr509, Thr514, and Ser517.**

While the hydroxyl group of Ser522 did not show significant differences, the hydroxyl group of Ser518 was significantly more surface exposed in the reduced conformations compared to the oxidized conformations. We thus propose that the phosphorylation of Ser522 by the priming kinase CDK5[8] may be independent of the redox state of CRMP2 (**Figure 8**). However, the subsequent phosphorylation steps by GSK3β[8] may require the protein to be in the reduce form, to allow phosphorylation of Ser518 followed by the other neighboring residues. Hence, the results presented here suggest a complex cross talk between the thiol switch and interconversion by phosphorylation in the semaphorin 3A signaling pathway.

## Acknowledgements

We thank Norman Geist (Greifswald) for continuous assistance, programming of some our analyzes tools and useful discussions during the simulations. CHL appreciates financial support of the Deutsche Forschungsgemeinschaft (DFG SPP 1710 (LI984/3-1)). The replica exchange MD simulations have been performed in part on the computers of the North-German Supercomputing Alliance (HLRN), project mvc00007.

# Tables

Table 1: Overview of CRMP2-simulations with color scheme for Figure 2, Figure 3, Figure 5, Figure 6 and SI 2, SI 3. Oxidation status, bivalent ion concentration, box dimensions and simulation time. Secondary structure elements: for X-ray data, MD simulation (with 2064 amino acids) of the oxidized CRMP2 (oxi) and reduced forms with different ion concentrations (red1 to red6) as an average over the last 25 ns (of every simulation) per element and for oxidized and reduced recombinant CRMP2 with CD-spectroscopy. Composition of the water exposed surfaces (SASA) for every simulation from hydrophobic amino acids to total SASA of the protein.

|  | oxidized | reduced | | | | | |
|---|---|---|---|---|---|---|---|
| Run: color scheme | oxi: black | red4: red | red1: blue | red2: green | red3: cyan | red5: orange | red6: grey |
| Bivalent ions | 1.63 mM $Ca^{2+}$ 4 $Ca^{2+}$-ions* | 0 mM no bivalent ions | 1.56 mM $Ca^{2+}$ 4 $Ca^{2+}$-ions* | 20 mM $Mg^{2+}$ 46 $Mg^{2+}$-ions | 200 mM $Mg^{2+}$ 459 $Mg^{2+}$-ions | 20 mM $Ca^{2+}$ 46 $Ca^{2+}$-ions | 200 mM $Ca^{2+}$ 459 $Ca^{2+}$-ions |
| Box-size in Å x | 159.0 | 155.8 | 159.3 | 159.8 | 155.1 | 155.8 | 155.3 |
| y | 156.6 | 167.5 | 175.1 | 167.5 | 166.7 | 167.5 | 167.0 |
| z | 178.8 | 160.3 | 166.3 | 160.3 | 159.6 | 160.3 | 159.8 |
| Water molecules | 136207 | 127468 | 142211 | 127438 | 126217 | 127434 | 126355 |
| Counter ions | 44 $Na^+$ | 52 $Na^+$ | 44 $Na^+$ | 40 $Cl^-$ | 866 $Cl^-$ | 40 $Cl^-$ | 866 $Cl^-$ |
| Simulation time | 280 ns | 275 ns | 260 ns | 260 ns | 260 ns | 285 ns | 220 ns |
| Secondary structure elements from MD during the last 25 ns | | | | | | | |
| Helix | 31.5 % | 32.9 % | 31.7 % | 33.4 % | 32.8 % | 33.2 % | 32.9 % |
| Turn | 26.1 % | 25.2 % | 27.1 % | 25.5 % | 25.7 % | 25.0 % | 25.2 % |
| Total helix and turn | 57.6 % | 58.1 % | 58.8 % | 58.9 % | 58.5 % | 58.2 % | 58.1 % |
| Sheet | 20.4 % | 20.3 % | 20.5 % | 20.2 % | 20.2 % | 20.9 % | 20.3 % |
| Coil | 22.0 % | 21.6 % | 20.8 % | 20.8 % | 21.3 % | 21.0 % | 21.7 % |
| Secondary structure content from CD spectroscopy of the recombinant protein | | | | | | | |
| Helical | 57.4 % | | | 62.9 % | | | |
| Sheet | 12.8 % | | | 10.4 % | | | |
| Random | 28.0 % | | | 26.5 % | | | |
| Secondary structure content in the X-ray data of reduced CRMP2 crystals | | | | | | | |
|  | | | | 2VM8 (1905 aa) | | 2GSE (1901 aa) | |
| Helix | | | | 37.5 % | | 37.7 % | |
| Turn | | | | 19.0 % | | 18.8 % | |
| Total helix and turn | | | | 56.5 % | | 56.5 % | |
| Sheet | | | | 23.3 % | | 24.3 % | |
| Coil | | | | 20.2 % | | 19.3 % | |
| Composition of water exposed surfaces during the last 25 ns of MD | | | | | | | |
| Hydrophobic aa | 29.1 % | 29.0 % | 29.4 % | 29.6 % | 29.7 % | 29.5 % | 29.6 % |

*concentration of the four (structure supporting) $Ca^{2+}$-ions in the simulation box for comparison to *in vitro* experimental concentrations